\documentclass[%
 reprint,
 amsmath,amssymb,
 aps,
onecolumn,prl]{revtex4-1}

\usepackage{graphicx}% Include figure files
\usepackage{epstopdf}
\usepackage{dcolumn}% Align table columns on decimal point
\usepackage{bm}% bold math
\usepackage{lineno}
\linenumbers
\begin{document}

\preprint{APS/123-QED}

\title{ Robust estimate of  dynamo thresholds in the von K\'arm\'an sodium experiment using the Extreme Value Theory}% Force line breaks with \\

\author{Davide Faranda,}
\affiliation{Laboratoire SPHYNX, Service de Physique de l'Etat Condens\'e, DSM, CEA Saclay, CNRS URA 2464, 91191 Gif-sur-Yvette, France}
 \email{davide.faranda@cea.fr}
\author{Mickael Bourgoin}
\altaffiliation{also at Laboratoire des Ecoulements G\'eophysiques et Industriels, CNRS \& Universit\'e Joseph Fourier, BP 53, 38041 Grenoble Cedex 9, France}
\affiliation{Laboratoire de Physique, \'Ecole Normale Sup\'erieure de Lyon,  CNRS \& Universit\'e de Lyon,\\ 46 all\'ee d'Italie, 69364 Lyon Cedex 07, France}
 \author{Sophie Miralles}
\affiliation{Laboratoire de Physique, \'Ecole Normale Sup\'erieure de Lyon,  CNRS \& Universit\'e de Lyon,\\ 46 all\'ee d'Italie, 69364 Lyon Cedex 07, France}
\author{Philippe Odier}
\affiliation{Laboratoire de Physique, \'Ecole Normale Sup\'erieure de Lyon,  CNRS \& Universit\'e de Lyon,\\ 46 all\'ee d'Italie, 69364 Lyon Cedex 07, France}
\author{Jean-Francois Pinton}
\affiliation{Laboratoire de Physique, \'Ecole Normale Sup\'erieure de Lyon,  CNRS \& Universit\'e de Lyon,\\ 46 all\'ee d'Italie, 69364 Lyon Cedex 07, France}
\author{Nicolas Plihon}
\affiliation{Laboratoire de Physique, \'Ecole Normale Sup\'erieure de Lyon,  CNRS \& Universit\'e de Lyon,\\ 46 all\'ee d'Italie, 69364 Lyon Cedex 07, France}
\author{Francois Daviaud}
\affiliation{%
Laboratoire SPHYNX, Service de Physique de l'Etat Condens\'e, DSM,
CEA Saclay, CNRS URA 2464, 91191 Gif-sur-Yvette, France}
\author{B\'ereng\`ere Dubrulle}
\affiliation{Laboratoire SPHYNX, Service de Physique de l'Etat Condens\'e, DSM, CEA Saclay, CNRS URA 2464, 91191 Gif-sur-Yvette, France}



\begin{abstract}
 We apply a new threshold detection method  based on the extreme value theory to the von K\'arm\'an sodium (VKS) experiment data. The VKS experiment is a successful attempt  to get  a dynamo magnetic field in a laboratory liquid-metal experiment. We first show that the dynamo threshold is associated to a change of the probability density function of the extreme values of the magnetic field. This method does not require the measurement of response functions from applied external perturbations, and thus provides  a simple threshold estimate. We apply our method to different configurations in the VKS experiment showing that it yields a robust indication of the dynamo threshold as well as evidence of hysteretic behaviors. Moreover, for the experimental configurations in which a dynamo transition is not observed, the method provides a way to extrapolate an interval of possible threshold values.

\end{abstract}

\pacs{Valid PACS appear here}% PACS, the Physics and Astronomy
\maketitle

%\tableofcontents

 It is generally accepted that the planetary magnetic field is generated by  dynamo action, an instability mechanism inside the liquid  conducting fluid of the planetary core. There is however presently no general theory providing an estimate for the corresponding dynamo threshold, except in some particular cases \cite{stieglitz2001experimental,radler1998karlsruhe,gailitis2001magnetic}. The main difficulties in computing the threshold derive from the turbulent nature of the flow, that make the dynamo action akin to a problem of instability in a presence of a multiplicative noise~\cite{leprovost2005turbulent}. As more and more data  from  experiments are available \cite{berhanu2007magnetic,spence2007turbulent,kelley2007inertial}, the possibility of devising precise, almost automated methods for dynamo threshold
 detection would be welcome. 
 The statistical approach to this question traditionally involves so-called indicators of criticality~\cite{scheffer2009early}. Some of these indicators are based on modifications of the auto-correlation properties of specific observables when parameters controlling the system approach some critical value, others on the fact that an increase of the variance and the skewness is observed when moving towards tipping points \cite{kuehn2011mathematical}. Other approaches are based on the definition of \textit{ad hoc} susceptibility functions or critical exponents \cite{monchaux2009Karman,berhanu2009bistability,miralles}. In \cite{lahjomri1993cylinder,miralles}, the decay of external applied magnetic field pulses is studied and the transition is detected through the divergence of the decay times near  the dynamo threshold.  Although interesting for controlled laboratory applications, this approach cannot be extended to problems involving planetary scales. 
 In the present paper, we suggest that  the statistical approach based on the Extreme Value Theory proposed in \cite{farandamanneville} could provide a robust  determination of the threshold even in the presence of turbulence. The main advantage of the present method is that it yelds to a precise and  unique determination of the threshold as the location of zero crossing of a statistical parameter $\kappa$. It therefore works even in the case of imperfect bifurcation that usually occurs in experimental dynamo due to the ambient magnetic field (Earth field,
residual magnetization of the disks and other magnetic perturbations of the set up). To illustrate the possibilities of the method, we analyse data from the VKS experiment, consisting  of a von K\'arm\'an swirling flow of liquid sodium. In this experiment, turbulent effects are roughly of the same order as the mean flow. The control parameter of the system is the magnetic Reynolds  number: $Rm$ which is proportional to the driving impellers rotation frequency $F$. Several dynamo and non dynamo configurations  have been  obtained by changing the material of the impellers and of the  cylinder \cite{miralles,boisson2012symmetry}  and by varying the impellers rotation frequency.  This versatility allows for reproducing a spectrum of magnetic field dynamics which can be observed for the planetary magnetic fields such as reversal \cite{berhanu2007magnetic}, bistability \cite{berhanu2009bistability,miralles2} or localization \cite{gallet2012experimental}.  Applying our method to several different configurations, we show in the present article that it provides a robust indication of the dynamo threshold  as well as  evidence of hysteretic behaviors. \\

\paragraph*{Method}
We use the statistical approach based on the Extreme Value Theory proposed in \cite{farandamanneville} as a criterion allowing the determination of the dynamo threshold. We briefly recall the basic intuition beyond the method referring to \cite{farandamanneville} for further discussions.
Classical Extreme Value Theory (EVT) states that, under general assumptions, the statistics of maxima $M_m=\max\{ X_0,X_1, ..., X_{m-1}\}$ of independent and identically distributed (i.i.d.) variables $X_0, X_1,\dots, X_{m-1}$, with cumulative distribution function (cdf) $F(x)$ in the form:
$$
F(x)=P\{a_m(M_m-b_m) \leq x\},
$$
where  $a_m$ and $b_m$ are normalizing sequences, asymptotically obeys  a Generalized Extreme Value (GEV) distribution with cumulative distribution function:
\begin{equation}
F_{G}(x; \mu, \sigma,
\kappa)=\exp\left\{-\left[1+{\kappa}\left(\frac{x-\mu}{\sigma}\right)\right]^{-1/{\kappa}}\right\}
\label{cumul}
\end{equation}
with $1+{\kappa}(x-\mu)/\sigma>0 $.  The {\it location parameter\/} $\mu \in \mathbb{R}$ and  the {\it scale parameter\/} $\sigma>0$ in Equation~\ref{cumul} account for the normalization of the data, avoiding the recourse to scaling constants  $a_m$ and $b_m$ \cite{LLR83}. 

The sign of $\kappa$ discriminates the kind of tail decay of the parent distribution:
When ${\kappa} = 0$,  the distribution is of Gumbel type (type 1). This is the asymptotic Extreme Value Law (EVL) to be expected when the parent distribution shows an exponentially decaying tail. The Fr\'echet distribution (type 2), with $\kappa>0$, is instead observed when the parent distribution possess  a fat tail decaying  as a power law. Eventually, the Weibull distribution (type 3), with $\kappa<0$, corresponds to a parent distribution having a finite upper endpoint. When properties of maxima and minima are of interest, respectively corresponding to the exploration of the right or left tails of the parent distribution, they can be treated on an equal footing by considering the minima as maxima of the variables after sign reversal \citep{coles2001introduction}. 
Physical observables have generally bounded fluctuations and   their extremes follow Weibull distributions~\cite{holland2012extreme,lucarini2012extreme}. Gaussian fluctuations (featuring Brownian motion of microscopic degrees of freedom) would yield the formal possibility of infinite extremes and thus Gumbel distributions, but the convergence towards this law is logarithmically slow \cite{hall1979rate} so that a Weibull law is observed in these cases as well.\

The interest of the EVL statistics in bifurcation detection relies on the change of the nature of the fluctuations
of a given system, when going from a situation with one stable attractor to a situation with two competing attractors, with jump between the two allowed either under the effect of external noise or due to internal chaotic fluctuations. In such a case, two time scale are present, a short one related to transitive dynamics within an attracting component and a long one corresponding to intermittent jumps from one to the other component. 
The fluctuations and their extreme are then of different nature over the two time scales: over the long time scale,  some extremes correspond to noisy excursions directed toward the saddle-state and gain a {\it global\/} status as they can trigger jumps from one to the other component. The probability increases as the observable visits corresponding ``anomalous'' values associated to these global extremes during a time series of length $s$, and the tail of the parent distribution becomes large. 
Through the bifurcation, we are thus in a situation where the parent distribution goes from bounded fluctuations (with extreme converging to a Weibull law) to fluctuations with fat tails (with extreme converging to a  Fr\'echet distribution). 
The shape parameter $\kappa$ then changes through the bifurcation from $\kappa<0$ to $\kappa>0$, wich enables a precise definition of the threshold as  the value at which the zero  crossing of $\kappa$ happens.  Physical observables will display deviations of greater amplitude in the direction of the state the system is doomed to tumble, than in the opposite direction, therefore one expects to observe this switching either in the maxima or in the minima.\\

\paragraph*{Experimental set-up}

\begin{figure}[ht]
\centering
\includegraphics[width=80mm]{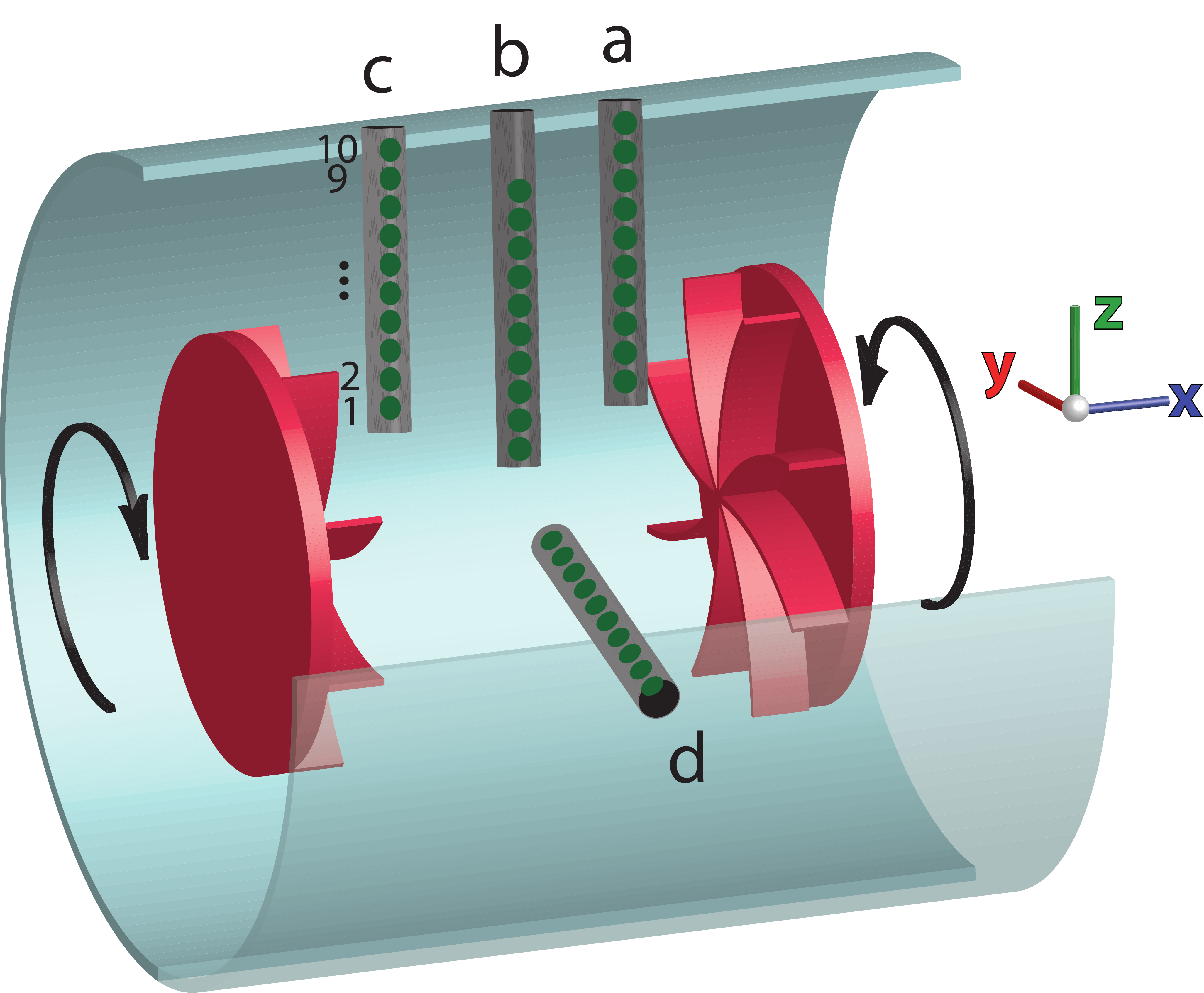}
\caption{Experimental setup, showing the location of the Hall probes. $x$ is the axial coordinate directed from impeller 1 to impeller 2 }
\label{manip}
\end{figure}

\begin{figure}[ht]
\centering
\includegraphics[width=140mm]{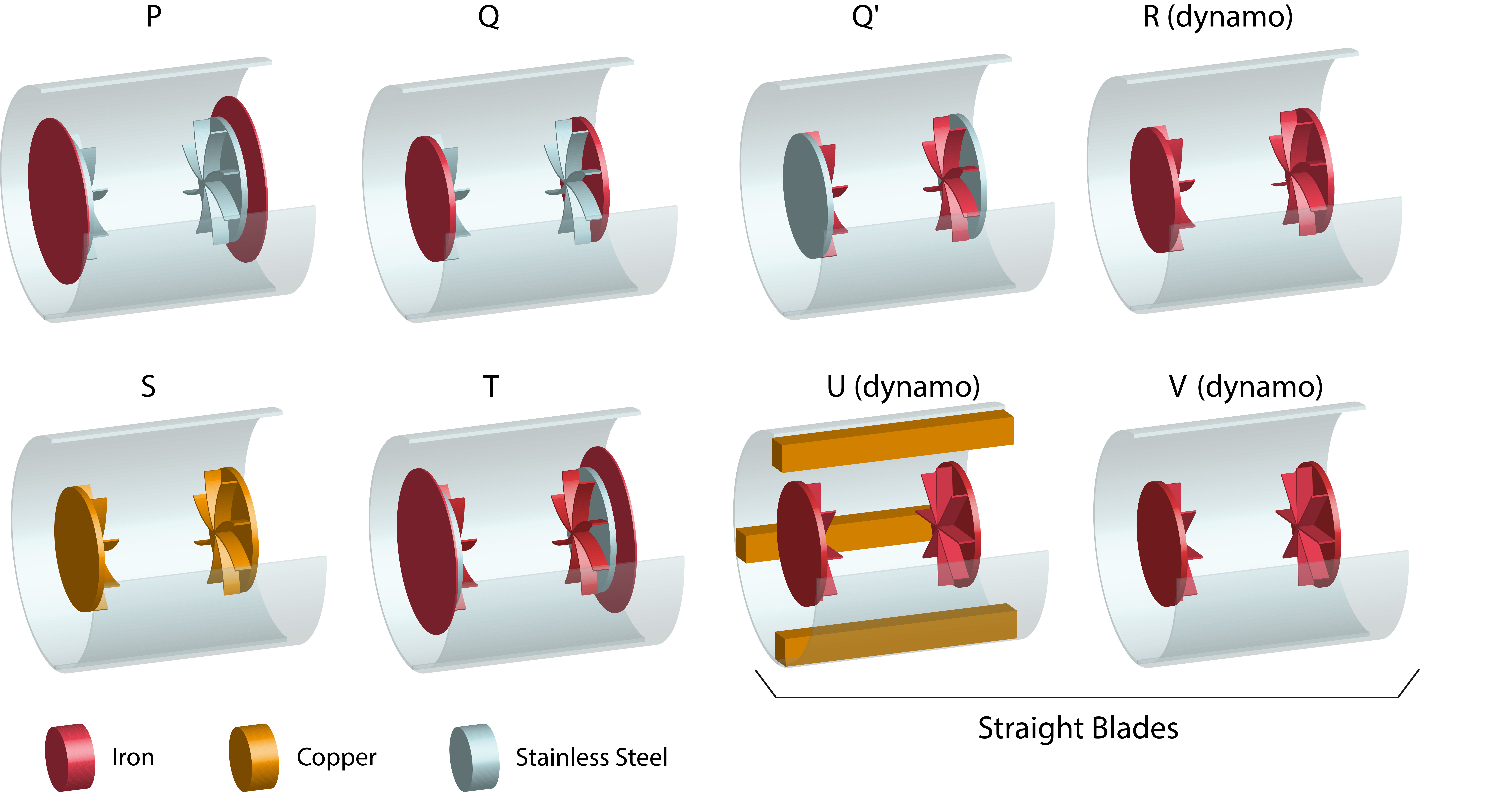}
\caption{Schematic representation of the studied VKS configurations. Gray colors stands for stainless steel, yellow color for copper and red for soft iron.}
\label{configurations}
\end{figure}

 Here we focus on  the VKS experiment, consisting  of a von K\'arm\'an swirling flow of liquid sodium. The dynamo is generated in a cylinder of radius $R_0=289$ mm by the motion of two coaxial discs of radius $R_{imp}=154.5$ mm, counter-rotating at a frequency $F$.  We define the   magnetic Reynolds number as: $Rm=2\pi\mu_0 \sigma R_{imp} R_0 F$ where $\sigma=9.6\times10^6$ $\Omega\cdot{\rm m}^{-1}$  is the sodium electrical conductivity and $\mu_0$ the permeability of vacuum. In the sequel, we use data from the 8 configurations obtained by changing the material of the impellers and of the  cylinder as shown  in Fig.~\ref{configurations} and described in \cite{miralles}.
Magnetic fields are recorded using four arrays of ten 3-axis Hall effect sensors inserted in radial shafts, as shown in Fig. \ref{manip}. Two arrays are inserted in the mid-plane of the vessel, within long probe shafts (labeled b and d in Fig.~\ref{manip}); the other two  are inserted closer to the impellers, within shorter probe shafts (labeled a and c in Fig.~\ref{manip}). These magnetic field at the sensors are recorded at a rate of 2000 Hz, with accuracy $\pm 0.1$~G.
Overall, the probes provide  measurements of the 3 components of the magnetc field $\vec{B}(t)$ as a function of time $t$.

\paragraph*{Application to VKS data.}
 We  present the results for   the  detection of the dynamo  threshold $Rm^*$  by using as observable the modulus of the magnetic field  $|\vec{B}(t)|$ measured by the 40 different detectors (Hall probes).

\begin{figure}[ht]
\centering
\includegraphics[width=80mm]{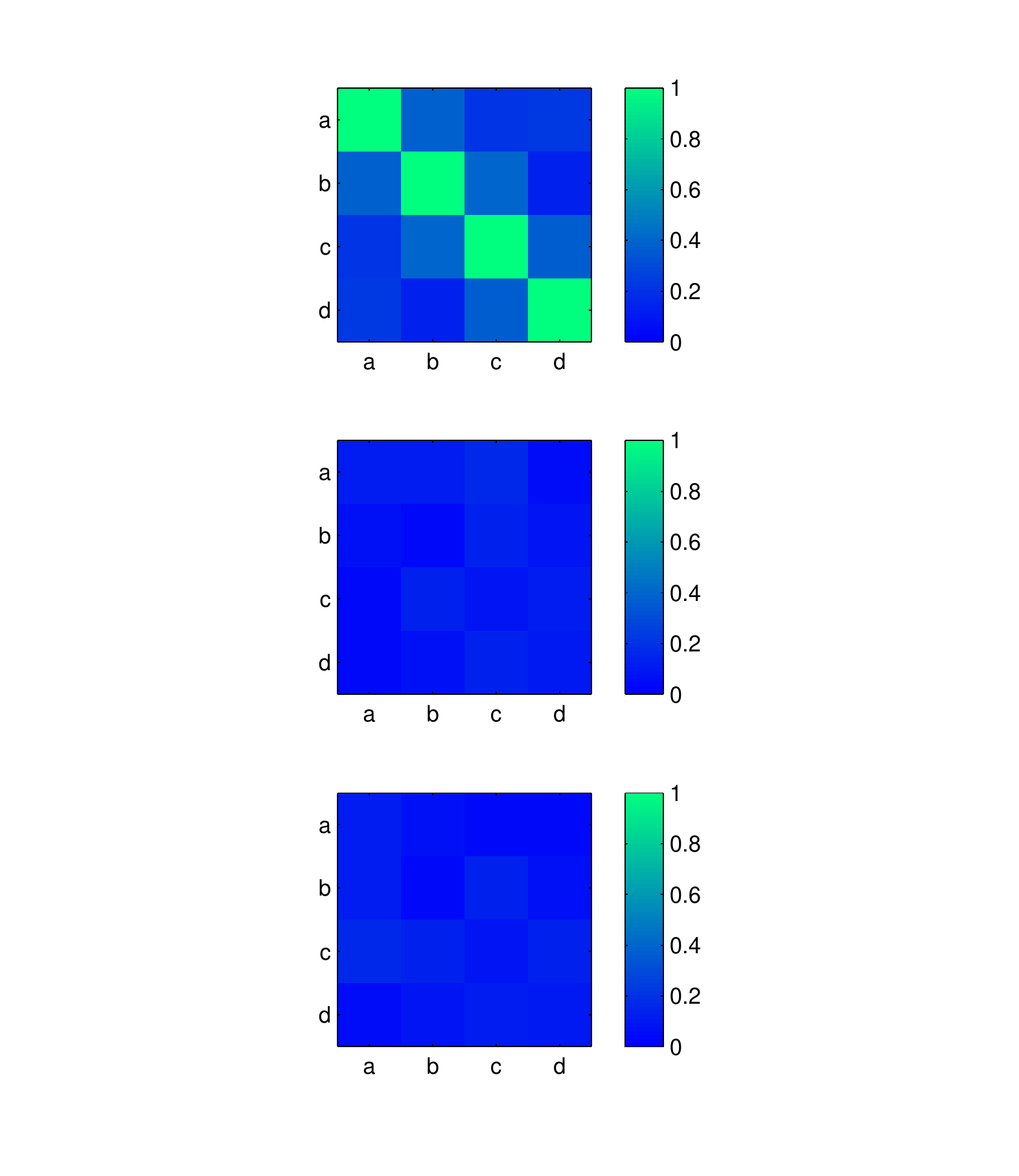}
\includegraphics[width=80mm]{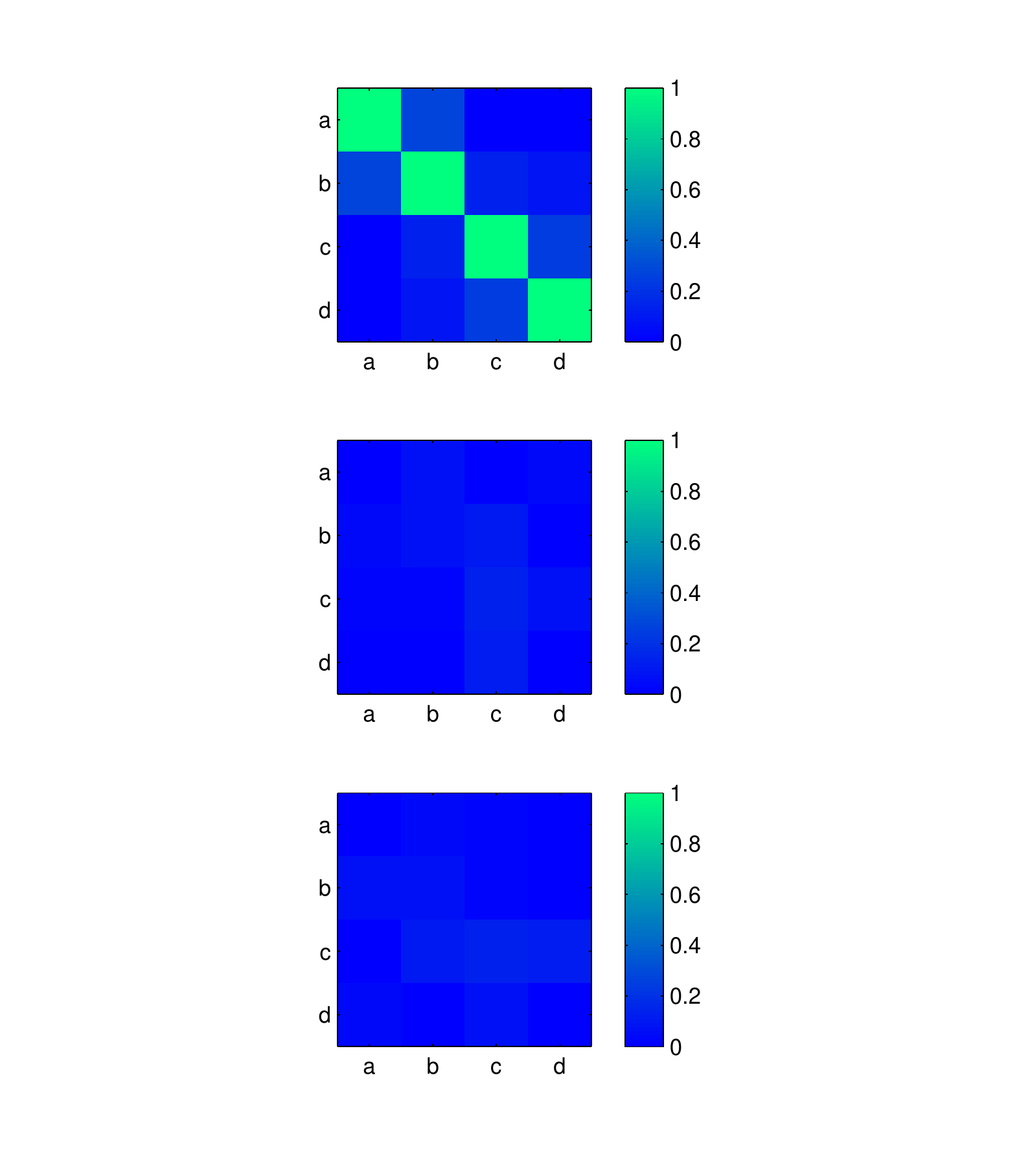}
\caption{ Cross correlation $\tilde{\tau}_{M(\lambda),M(\mu)}(h)$  for $m=1000$ (left panels) and $m=4000$ (right panels) $\lambda=\{a,b,c,d\}$, $\mu=\{a,b,c,d\}$. From top to bottom panels: $h=0$, $h=5$, $h=-5$. R configuration, $Rm\simeq 33.8$ }
\label{crosscorr}
\end{figure}

The method can be described as follows. First of all, the extremes of the magnetic field  are extracted by using the  so called  {\em block maxima approach} which consists in dividing the series  $|\vec{B}(t)|, \ t=1,2,...,s$  into $n$ bins each containing $m$ observations ($s=nm$) and thus selecting the maximum (minimum) $M_j$ in each bin. The series of $M_j, j=1,...,n$ is then fitted to the  GEV distribution via the L-moment procedure described in \cite{faranda2012generalized}.  
In order to sample proper extreme values one has to consider a bin length longer than the correlation time $ \tau $.  
For each of the   sensors   we have  computed  $\tau$  as the first zero of the autocorrelation function finding that $0.42$ s $< \tau <$ $1$ s lags depending on  the cases considered. This value is similar to the magnetic diffusion time found in \cite{bourgoin2002magnetohydrodynamics}. 
By choosing   a bin duration longer than $1$ s (or, equivalently, a number of samples $m$ in each bin larger than 2000)  and repeating the fit until the shape parameter $\kappa$ is not changing in appreciable way, one can establish  the convergence to the GEV model   \cite{LLR83}.  
In our experiments we found that reliable estimates can be  generally obtained for  $m>4000$.  Being the length of each series  $ 10^5<s<3\cdot10^5$, for any choice  of $m>4000$  no more than $n=100$ maxima can be extracted. Such a value of  $n$ is  one order of magnitude smaller than  the one prescribed in \cite{eckmann1992fundamental,faranda2011numerical} for avoiding biased fits to the GEV model.  In order to overcome this problem  we have   grouped  sensors  located at the same radial position. The sensors of the four arrays of Hall effect sensors are not installed at the same radial distance (see Fig~\ref{manip} for a visual explanation). However, an effective radial grouping can be obtained    by adding to the $n$ extremes of the sensor $a_l$    the ones of $b_{l+2},c_l$ and $d_{l+2},\ l=1,...,8$, thus obtaining 8 different series with a sufficient number of maxima to perform the fit.
The  choice of grouping the sensors  by their radial location is justified by checking that the shape of the distribution, which enters in the computation of the shape parameter $\kappa$ does not change substantially for sensors located at the same radial position. In order to do so, we have computed the skewness and the kurtosis for the time series of the magnetic field, finding small variations  for sensors located at the same radial position.  
We also checked that the maxima extracted by combining the series are independent   by analyzing  the   cross-correlation function of different sensors. For example, for sensors $a$ and $b$, the cross-correlation function is defined as:
$$\tilde{\tau}_{M(a),M(b)}(h)=\frac{1}{n}\sum_{i=1}^{n-h} (M_j(a)- \langle M_j(a)\rangle_j)(M_{j+h}(b)- \langle  M_j(b)\rangle_j).$$ Here, the notation $\langle \cdot\rangle_j$ indicates the expectation value taken over the $j$ index.  The results of this analysis are shown in Fig.~\ref{crosscorr}, for maxima in the case $Rm\simeq33.8$, sensors index 5, in the R configuration. The plots on the left refer  to $m=1000$, the ones on the right to $m=4000$.  From top to bottom we represent $\tilde{\tau}(h)$ respectively for $h=0$, $h=+5$ and $h=-5$. The case $h$=0 corresponds to sensors located at the same radial position, which we grouped in our study. One can observe that although the correlation is non-zero, it is relatively small (about 0.5 for neighboring probes and smaller than 0.2 for non-neighbouring probes), which validates our grouping of the sensors to increase our statistics. In addition, for sensors located at different radial positions the decorrelation is total (see the example at $h=\pm 5$ in figure 3 but the decorrelation already starts at $h=\pm 1$). This indicates that the different series we  show are totally independent.\\

Due to the different size of the fluctuations, extremes have been renormalized  using the following, rather standard, definition:
$$\tilde{M}_j(a_l) =\frac{(M_j,a_l -  \langle M(a_l)\rangle_j)}{ \sqrt{\langle M_j(a_l) - \langle M(a_l)\rangle_j \rangle_j}}$$

The same normalization applies  for the sensors $b,c,d$.   There are less trivial ways of normalizing the extremes e.g. by choosing other location indicators than the expected value such as the median or the mode (the most probable value). We thus tested that by replacing the mean with such indicators and checked that  the results do not change in an appreciable way. 

\begin{figure}[ht]
\centering
\includegraphics[width=130mm]{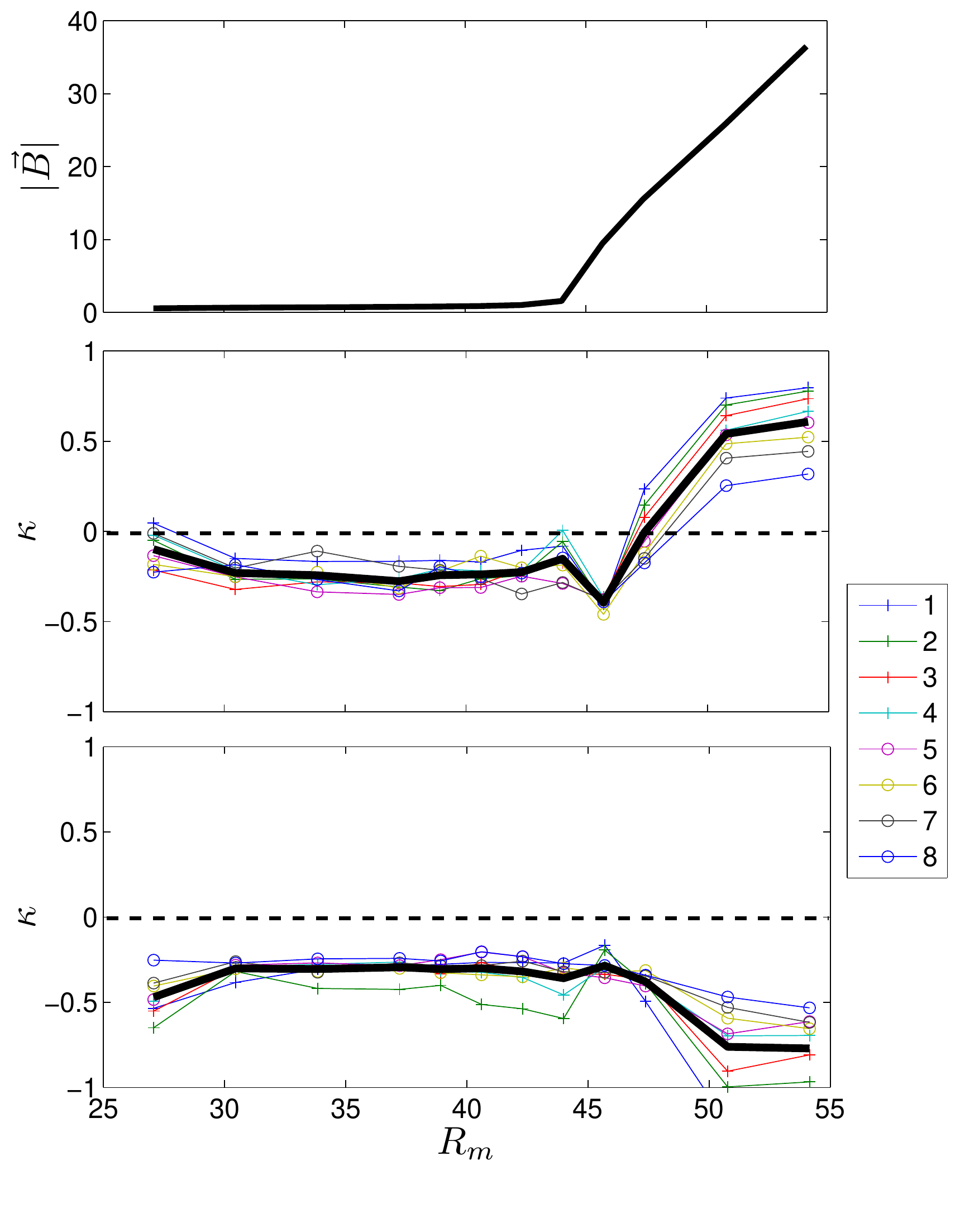}
\caption{Upper panel: bifurcation in terms of the  magnetic field averaged over all the sensors. Central panel: Shape parameters vs Reynolds magnetic number for the 8 group of sensors (each in a different color), $<k>_l$ , (thick black line) and Gumbel law $\kappa=0$ (dashed line) in the R configuration.  Lower panel: same as the central panel but   for the minima.}
\label{mm}
\end{figure}

\begin{figure}[ht]
\centering
\includegraphics[width=80mm]{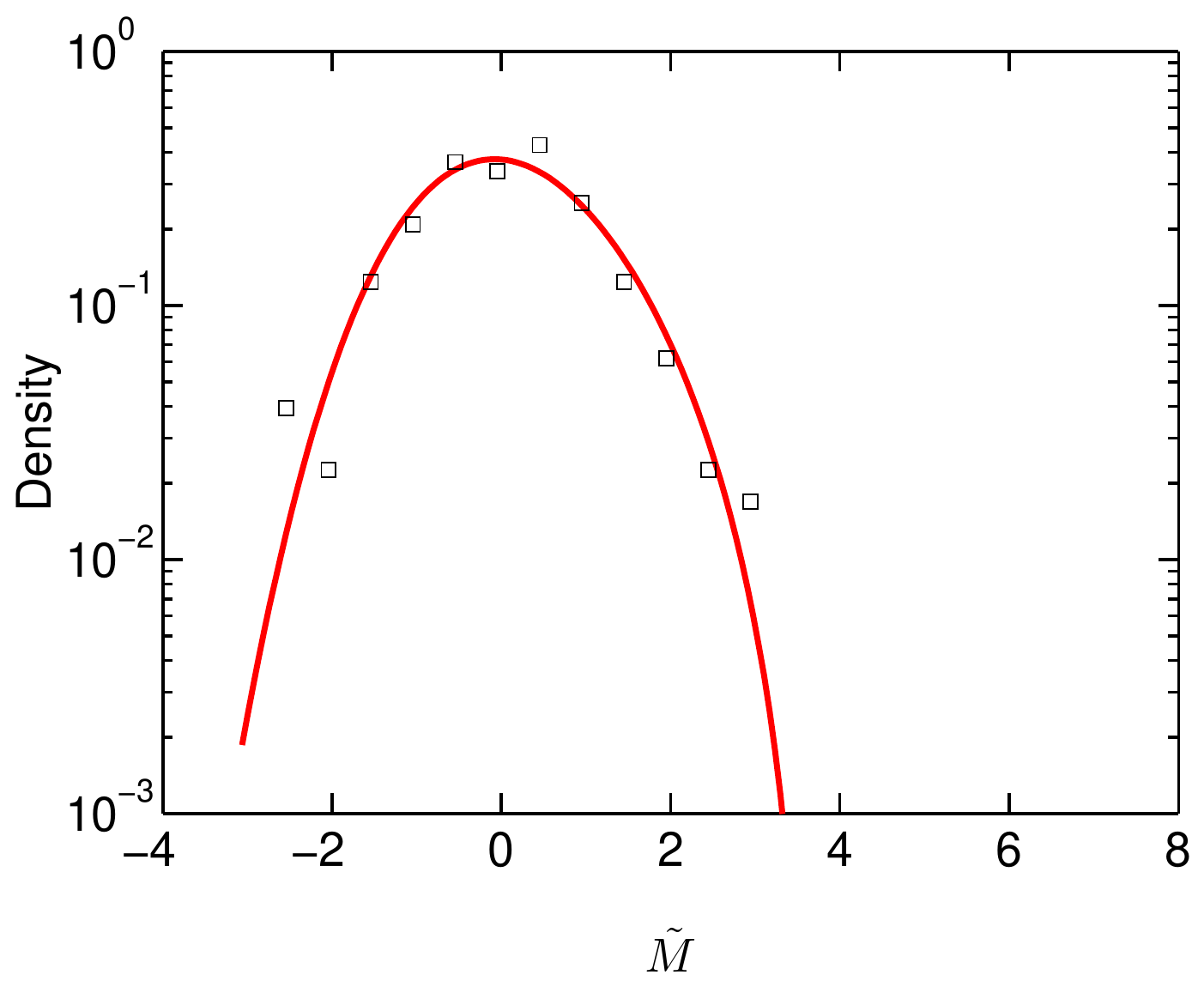}
\includegraphics[width=80mm]{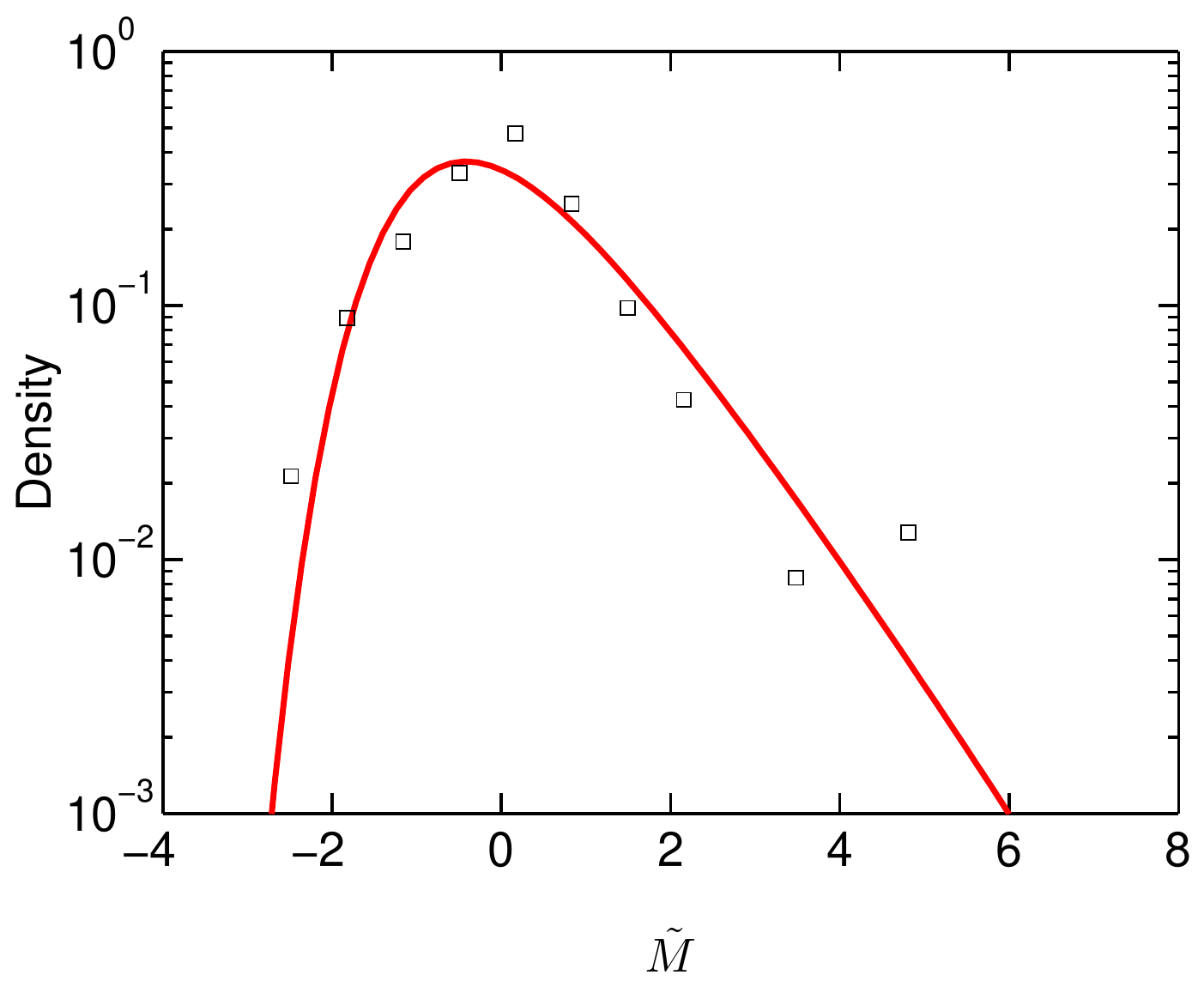}
\caption{Two histograms for the normalized maxima  $\tilde{M}$ of the sensors 6 (black markers) and correspondent fits to the GEV distribution (red lines). Left: $Rm\simeq27$ the maxima are bounded: $\kappa=-0.21$. Right: $Rm\simeq48$, some maxima are detached with respect to the bulk statistics. These events trigger the transition of $\kappa$ towards positive values: $\kappa=+0.01$.}
\label{histo}
\end{figure}

\begin{figure}[ht]
\centering
\includegraphics[width=100mm]{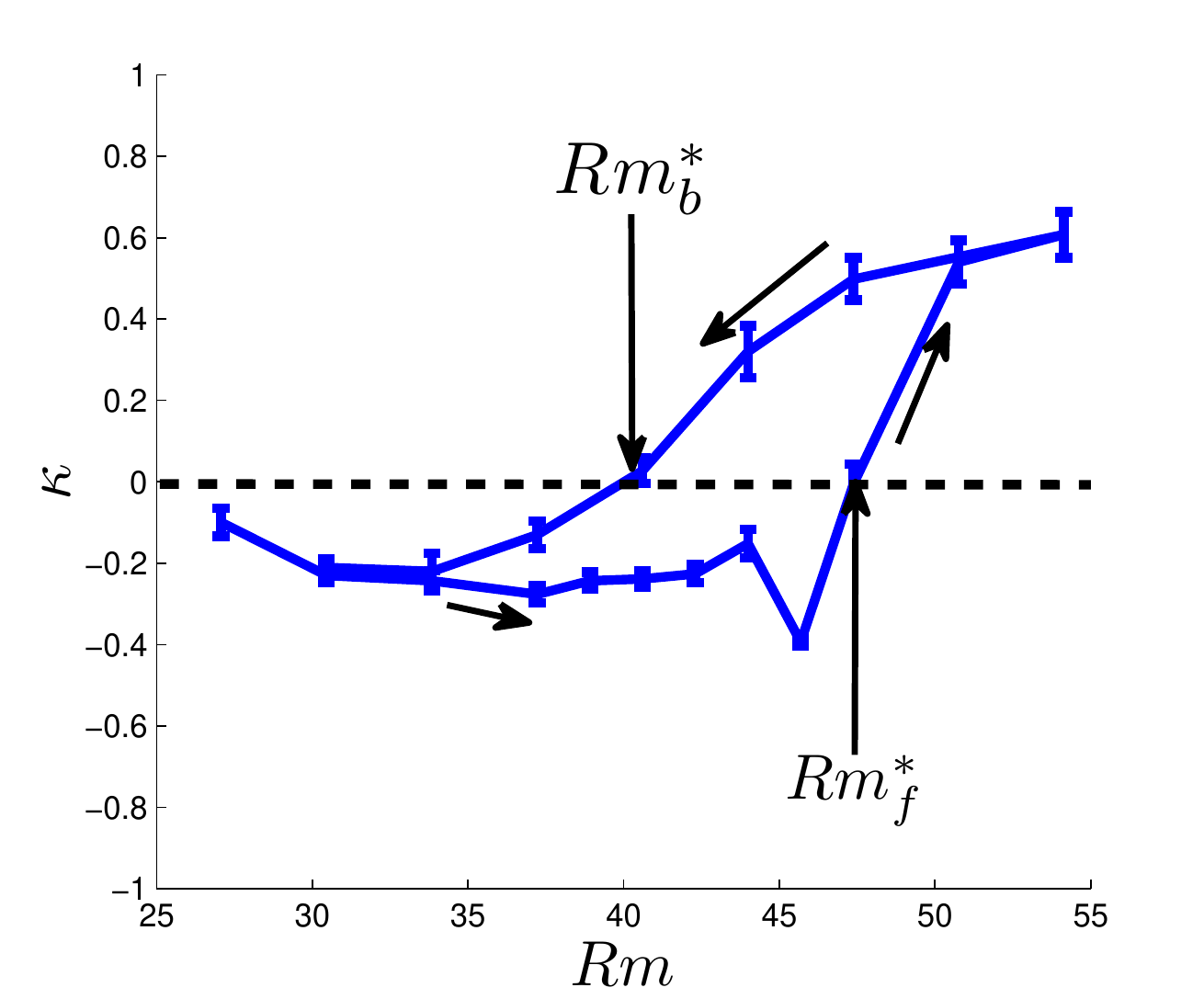}
\caption{$\langle \kappa \rangle_l$, (blue line) and Gumbel law $\kappa=0$ (dashed black line)  vs magnetic Reynolds  number in the R configuration. The  arrows indicate the direction of variation of $Rm$ in the experiment. }
\label{hyster}
\end{figure}

\paragraph*{Results.}
We begin the analysis by computing dynamo threshold $Rm^*$ in the experiments performed with the configuration R featuring soft iron impellers. This configuration produces a well-documented stationary dynamo at $Rm\approx 44$, thereby providing a fair test of our method \cite{berhanu2010dynamo}.
In the run we analyze, the Reynolds magnetic number is increased monotonically from $Rm\simeq26$ up to $Rm\simeq54$. 
By monitoring the value of $|\vec{B}|$ as a function of $Rm$, represented in the upper panel of Fig.~\ref{mm}, one observes a sudden increase of the magnetic field amplitude around the value $Rm\approx 44$, leading to previous definition of the threshold
parameter as $Rm^*= 44$.\

The observation of variations of $|\vec{B}|$ provides interesting information about the detection of threshold through EVL method. Indeed, since  beyond the dynamo threshold  $Rm^*$ the values of $|\vec{B}|$ are significantly higher,  we expect to detect the transition by the change of sign of the maxima distribution whereas the minima shape parameter should remain negative even across the transition. Results are shown in Fig.~\ref{mm}  for the shape parameter of the maxima (central panel) and of the minima (lower panel). Each color represents the curve of $\kappa$ obtained by  grouping the sensors located at the same radial position  whereas the thick lines respectively  represent an average over $l$ (solid black line) and the Gumbel law (dashed black line). When $Rm$ is approaching 47   the average shape parameter for the distribution of maxima first decreases, then increases and changes sign at $Rm=47$, whereas for the minima it remains negative. We therefore set the threshold value  $Rm^*\simeq47$.  The decrease before the change of sign might be a signature of earth-field expulsion before dynamo onset. The change of sign, characteristic of dynamo onset, is associated with a change in the nature of distribution of maxima of the magnetic field, as expected from EVL theory. Indeed, we have plotted in Fig.~\ref{histo} two histograms for the maxima distribution, one for a value of $Rm$ far from the transition (left plot) and 
one for $Rm$ close to the bifurcation (right plot). Whereas in the first case the distribution of maxima is  bounded above, in the second case the largest values of $\tilde{M}$ will eventually trigger the transition and are responsible for the change of sign of $\kappa$.  One may note in Fig.~\ref{mm} that, contrary to the transition presented in \cite{farandamanneville}, here there is an evident effect also on the minima shape parameter which tends to more negative values for  $Rm>Rm^*$. This effect, definitely due to the complex geometry of the two attracting basins involved in the transition, is difficult to quantify and will be addressed specifically in future publications.

By analyzing the results obtained  at low frequencies of rotations, the fit for each  group of sensors returns a shape parameter  statistically dispersed around the average, with no radial dependence. On the contrary, for $Rm>Rm^*$ the shape parameter crosses zero  for increasing values of $Rm$ as the radial location of the sensors increases. This effect is even more pronounced for sensors outside the flow (i;e; sensor 9 and 10 of probes a and c, not shown in Fig.~\ref{mm} since at  these radial locations only two sensors were available instead of 4).  This means that the threshold detection based only on external sensors is likely to overestimate the threshold. This has, of course, great implications for the detection of threshold of magnetic fields from planetary observation as we are likely to observe only an equivalent of the outer sensors.  This analysis confirms nevertheless a posteriori the reasonableness of grouping the sensors in a radial direction.\\

Hysteresis has been previously reported in the VKS experiment \cite{monchaux2009Karman,berhanu2010dynamo} and was also observed in the R configuration under scrutiny here: in order to shut down the dynamo one has to decrease the magnetic Reynolds number  to values smaller  than $Rm^*$. This is presumably an effect of the residual magnetization of the iron impellers. This hysteresis is a  good test for further validation of the results obtained via the extreme value based technique since the curve of the shape parameter should be able to detect some hysteretic behavior. If we redefine the dynamo activation threshold found in the previous analysis  as $Rm^*_f=47$, $f$ indicating the first passage in the forward direction of the experiment, we expect to find a dynamo deactivation threshold $Rm^*_b< Rm^*_f $,  $b$ indicating the backward experiment obtained by decreasing $Rm$ from  $Rm\simeq 55$ to $Rm=30$.  We have then analyzed a run in which the magnetic Reynolds number is first increased monotonically from $Rm\simeq26$ up to $Rm\simeq54$, then decreased monotonically from $Rm\simeq54$ up to $Rm\simeq26$.
 The results shown in Fig.~\ref{hyster} for  the maxima average shape parameter  $\langle \kappa \rangle_l\ l=1,...8$, clearly indicate the presence of a hysteresis cycle in agreement with expectation.   We have already commented on the forward part of the experiment repeated in Fig.~\ref{hyster} for clarity and represented by the right arrows.  When the frequency is instead decreased, a Fr\'echet extreme value law is observed until   $Rm^*_b\simeq 37 <Rm^*_f$. At this value, the shape parameter crosses the Gumbel law and  approaches again the Weibull distribution of the maxima. Note also that the  shape parameter for the minima (not shown here) remains always negative even in the backward transition, as expected by the theory described so far.

\begin{figure}[ht]
\centering
\includegraphics[width=100mm]{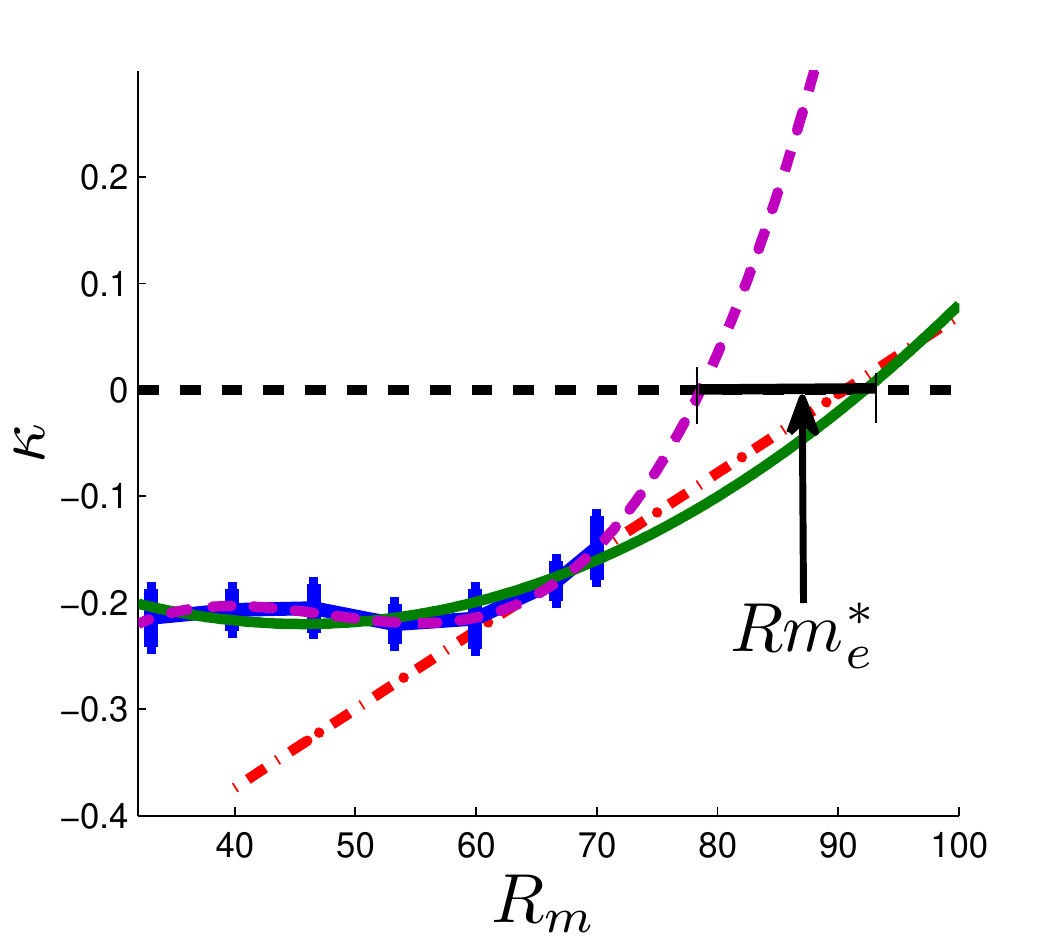}
\caption{$\langle \kappa \rangle_l$, (blue solid error-bar) and Gumbel law $\kappa=0$ (dashed black line) in the Q' configuration vs Reynolds magnetic number. The  red dashed-dotted line, the green solid line and  the magenta dashed line represent respectively a linear, quadratic and cubic fits of the data. The linear fit is obtained by considering only the   3 values of $\langle \kappa \rangle_l$ at higher $Rm$. $m=4000$ }
\label{extra}
\end{figure}

The same analysis has been carried out  for all the configurations shown in Fig.~\ref{configurations}.  The corresponding $Rm_f^*$  and $Rm^*_b$ are reported in the table below. For comparison, we have included in the table values estimated via three other techniques:  from the increase of the magnetic field amplitude $|B|$-denoted $Rm_{|B|}$- \cite{monchaux2009Karman,berhanu2009bistability},  
from decay time divergence $Rm^{d}$\cite{miralles} and via induction $Rm^i$ \cite{miralles}.
The value of the shape parameter  remains negative for both the maxima and the minima, in   the configurations P, Q, Q', S, T where dynamos have not been observed,  whereas the method is able to detect the dynamo and the hysteretic behavior for the U and V setup. These results are in agreement with \cite{miralles}.\
 For the configurations for which the dynamo (run P, Q, Q', S, T) is not observed within the range of accessible $Rm$, 
 it is interesting to follow~\cite{miralles}, and try to estimate possible dynamo threshold by extrapolation techniques.
 Indeed, in the Q', S and T, we observed that the values of  $\kappa$  increase monotonically for at least the 3 consecutive highest $Rm$. An example is shown for the Q' configuration  in Fig.~\ref{extra}. An extrapolate threshold value   $Rm^e$ can then be found by applying a polynomial fit of the $\langle \kappa \rangle_l$  curve and detecting the location of the zero crossing.   Of course, as seen in Fig.~\ref{extra}, the value of $Rm^e$ depends on the order of the polynomial fit: for example,   the value of $Rm^e$ obtained by a  linear and a quadratic fit is larger than what is obtained through   higher polynomial order fits. We then turned back to configurations R, U, V, and found that a cubic  fit of  the $\langle \kappa \rangle_l$ values such that $Rm< Rm^*$ provides an extrapolated threshold value $Rm^e$ that is close to  the $ Rm^*$ determined via real data. We thus run this cubic extrapolation technique to Q', S and T, and obtain value of $Rm^e$
 that are reported in the table. The extrapolated values  found here are generally smaller than the one found by Miralles et al. \cite{miralles}, but in both cases the extrapolation presents great uncertainty.\\

\begin{table}
\begin{center}
    \begin{tabular}{ | c |c | |c |c | c | | c | c |}
    \hline
    Run & $Rm_{||\vec{B}|}$ & $Rm^*_f$ & $Rm^*_b$ & $Rm^e$  & $Rm^d$  & $Rm^i$ \\ \hline
    P & - & - & - & - & - & -  \\ \hline
   Q & - & - & - & -  & - & 200 \\ \hline	
    Q' & - & - & - & 85$\pm$ 10   & 350 & 125 \\ \hline
    R & 44 & 46 & 37  & - &  51 & 56 \\ \hline
    S & - & - & - &  150$\pm$ 25  & - & - \\ \hline
    T & - & - & - & 100$\pm$ 25 & 250 & 205 \\ \hline
    U & 70 & 75 & 66  & - & 58 & 100 \\ \hline
    V & 66 & 67 & 45 & - & 71 & 93 \\ \hline
     \end{tabular}
     \caption{Dynamo threshold for various configuration in the VKS experiment, obtained through various technique:  $Rm_{||\vec{B}|}$: from the increase of the magnetic field amplitude $|\vec{B}|$  \cite{monchaux2009Karman,berhanu2009bistability}, $Rm^*_f$ and  $Rm^*_b$: forward and backward threshold obtained from the extreme value technique, with zero crossing detection (this paper);  $Rm^e$: from the extreme value technique, with cubic extrapolation to detect zero crossing (this paper); $Rm^{d}$: from decay time divergency extrapolation~\cite{miralles}; $Rm^i$  from induction increase extrapolation.}
\end{center}
\end{table}

In this article, we have tested a methodology for the detection of dynamo threshold based on EVT using datasets produced in the VKS experiment. This technique, applied here for the first time to an experimental dataset,  confirms the theoretical expectations of \cite{farandamanneville} and allows for detecting hysteretic behaviors. The main advantage of the technique is to provide a precise and unambiguous estimate of the thresholds on   probabilistic basis, providing the direction of the shift (towards the maxima or the minima). The analysis is affordable with every home PC and many software packages contain the routine necessary for performing the fit of the GEV distribution. In light of the possibility of extracting the magnetic field data from exoplanetary radio emissions, one could exploit the technique described in this article for studying the properties of exo-planetary magneto-spheres thus defining a criterion for the classification of planetary dynamos based on the detected threshold values. Moreover, since hysteretic behaviours are encountered in many other scientific fields, e.g the thermohaline circulation reversibility \cite{rahmstorf2005thermohaline}  in climate sciences  or the economical crisis behavior \cite{martin2012regional},  we consider the method  to be applicable to a more general class of problems featuring critical transitions. 

\section{Acknowledgments}
We thank the other members of the VKS collaboration, with whom the experimental runs have been performed. We thank M. Moulin, C. Gasquet, A. Skiara, N. Bonnefoy, D. Courtiade, J.-F. Point, P. Metz, V. Padilla, and M. Tanase for their technical assistance. This work is supported by ANR 08-0039-02, Direction des Sciences de la Mati\`ere, and Direction de l'Energie Nucl\'eaire of CEA, Minist\`ere de la Recherche, and CNRS. The experiment is operated at CEA/Cadarache DEN/DTN.

\bibliography{dynamo}% Produces the bibliography via BibTeX.

\end{document}